\documentclass[11pt]{article}
\usepackage{amsmath,amssymb,amsthm}
\usepackage{graphicx}
\usepackage{array}

\usepackage{multicol}
\usepackage[]{color}

\usepackage{algorithm}

\usepackage{algpseudocode}
\usepackage{array}

\newtheorem{prop}{Proposition}

\newtheorem{rem}{Remark}

\newcommand{\ssum}{\displaystyle\sum}

\newcommand{\F}{\mathbb F}

\begin{document}


\title{\Large \bf An Improvement of the
Cipolla-Lehmer Type Algorithms }

  \author{Namhun Koo$^1$, Gook Hwa Cho$^2$, Byeonghwan Go$^2$,  and Soonhak Kwon$^2$\\
  \small{\texttt{ Email: nhkoo@nims.re.kr, achimheasal@nate.com, kobhh@skku.edu, shkwon@skku.edu}}\\
  \small{National Institute for Mathematical Sciences, Daejeon, Republic of Korea$^1$}\\
  \small{Sungkyunkwan University, Suwon, Republic of Korea$^2$}
       }
\date{}



\maketitle

\begin{abstract}
Let $\F_q$ be a finite field with $q$ elements with prime power $q$
 and let $r>1$ be an integer with $q\equiv 1 \pmod{r}$.
In this paper, we present a refinement of the Cipolla-Lehmer type
algorithm given by H. C. Williams, and subsequently improved by K.
S. Williams and K. Hardy. For a given $r$-th power residue $c\in \F_q$
where $r$ is an odd prime, the  algorithm of H. C. Williams
determines a solution of $X^r=c$  in $O(r^3\log q)$ multiplications
in $\F_q$, and the algorithm of K. S. Williams and K. Hardy
finds a solution in $O(r^4+r^2\log q)$ multiplications in
$\F_q$. Our refinement finds a solution in $O(r^3+r^2\log q)$
multiplications in $\F_q$. Therefore our new method is better than
the previously proposed algorithms independent of the size of $r$,
and the implementation result via SAGE shows a substantial
speed-up compared with the existing algorithms.




\end{abstract}


\section{Introduction}

Let $r>1$ be an integer and $q$ be a power of a prime.  Finding
$r$-th root (or finding a root of $X^r=c$) in finite field
$\mathbb{F}_{q}$ has many applications in computational number
theory and in many other related topics. Some such examples
include point halving and point compression on elliptic curves
\cite{NIST}, where
 square root computations are needed. Similar applications for high
genus curves require $r$-th root computations also.

 Among several available root extraction methods of the equation $X^r-c=0$, there are
 two well known algorithms applicable for arbitrary integer $r>1$; the Adleman-Manders-Miller
 algorithm \cite{A.M.M},  a straightforward generalization of the
Tonelli-Shanks square root algorithm \cite{Shan,Ton} to the case
of $r$-th root extraction, and the Cipolla-Lehmer  algorithms \cite{Cip,Leh}.
Due to the cumbersome extension field arithmetic needed for the
Cipolla-Lehmer algorithm, one usually prefers the Tonelli-Shanks
or the Adleman-Manders-Miller, and  other related researches
\cite{A92, Bern, KCYL} exist to  improve the Tonelli-Shanks.

The efficiency of the Adleman-Manders-Miller algorithm heavily
depends on the exponent $\nu$ of $r$ satisfying $r^\nu | q-1$ and
$r^{\nu+1} \nmid  q-1$, which becomes quite slow if $\nu \approx
\log q$. Even in the case of $r=2$, it had been observed in
\cite{Muller} that, for a prime $p=9\times 2^{3354}+1$, running
the Tonelli-Shanks algorithm using various software such as Magma,
Mathematica and Maple cost roughly 5 minutes, 45 minutes, 390
minutes, respectively while the Cipolla-Lehmer costs under 1
minute in any of the above softwares. It should be mentioned that
such extreme cases (of $p$ with $p-1$ divisible by high powers of
$2$) may happen in some cryptographic applications. For example,
one of the NIST suggested curve \cite{NIST} P-224 $: y^2=x^3-3x+b$
over $\mathbb F_p$ uses a prime $p=2^{224}-2^{96}+1$.



A generalization to $r$-th root extraction of the Cipolla-Lehmer
square root algorithm is proposed by H. C. Williams \cite{HC} and
the complexity of the proposed algorithm is $O(r^3 \log q)$
multiplications in $\F_q$. A refinement of the algorithm in
\cite{HC} was given by K. S. Williams and K. Hardy \cite{KS} where
the complexity is reduced to $O(r^4+r^2 \log q)$ multiplications
in $\F_q$.  For the case of the square root,  a new Cipolla-Lehmer
type algorithm based on the Lucas sequence was given by M\"{u}ller
\cite{Muller}. A similar result for the case $r=3$ was also
obtained by Cho et al. \cite{Cho}, and a possible generalization
to the $r$-th root extraction of  M\"{u}ller's square root
algorithm was given in \cite{Cho2}.


In this paper,  we present a new Cipolla-Lehmer type algorithm for
$r$-th root extractions in $\F_q$ whose complexity is
$O(r^3+r^2\log q)$ multiplications in $\F_q$, which improves
previously proposed results in \cite{HC,KS}. We also compare our
algorithm with those in \cite{HC, KS} using the software SAGE, and
show that our algorithm performs consistently better than those in
\cite{HC,KS} as is expected from the theoretical complexity
estimation. In \cite{HC} and \cite{KS}, only the case where $r$ is
an odd prime was considered but we will give the general arguments
(i.e., no restriction on $r$) here.

The remainder of this paper is organized as follows: In Section 2,
we briefly summarize the Cipolla-Lehmer algorithm, and introduce
the works of H. C. Williams \cite{HC} and K. S. Williams and K.
Hardy \cite{KS}.  In Section 3, we present our refinement of the
Cipolla-Lehmer algorithm.   In Section 4, we give the  complexity
analysis of our algorithm and show the result of SAGE
implementations of the three algorithms (in \cite{HC}, \cite{KS},
and ours). Finally, in Section 5, we give the concluding remarks.

\section{Cipolla-Lehmer Algorithm in  $\mathbb{F}_{q}$}

Let $q$ be a prime power and $\F_q $ be a finite field with $q$
elements.  Let  $c\neq 0 \in \F_q$ be an $r$-th power residue in
$\F_q$ for an integer $r>1$ with $q\equiv 1 \pmod{r}$. We restrict
$r$ as an odd prime in this section.





\subsection{H. C. Williams' algorithm}

Let $b\in \F_q$ be an element such that $b^r-c$ is not an $r$-th
power residue in $\F_q$. Such $b$ can be found after $r$ random
trials of $b$. (See pp.479-480 in \cite{KS} for further
explanation.) Then the polynomial $X^r-(b^r-c)$ is irreducible
over $\F_q$ and there exists $\theta \in \F_{q^r}-\F_q$ such that
$\theta^r=b^r-c$.  Let $\omega=\theta^{q-1}=\left(b^r-c
\right)^\frac{q-1}r$. Then we have $\omega^r=1$ where $\omega$ is
a primitive $r$-th root because $b^r-c$ is not an $r$-th power in
$\F_q$.

 For all $0\leq i\leq r-1$, using $q\equiv 1 \pmod{r}$, one has
 $\theta^{q^i}=\theta\cdot
 \theta^{q^i-1}=\theta\cdot\left(\theta^{q-1}\right)^{1+q+\cdots+q^{i-1}}=\theta\omega^i$,
which implies $(b-\theta)^{q^i}=b-\theta^{q^i}=b-\omega^i \theta$.
Letting $\alpha=b-\theta$, one has
\begin{align}\label{Viete}
\alpha^{\sum_{j=0}^{r-1} q^j}&=(b-\theta)^{1+q+q^2+\cdots
+q^{r-1}}
 =\prod_{i=0}^{r-1}(b-\omega^i\theta)
 =b^r-\theta^r=c.
\end{align}

\noindent Thus one may find an $r$-th root of $c$ by computing
$\alpha^\frac{\sum_{j=0}^{r-1}q^j}{r} \in
\F_q[\theta]=\F_q[X]/\langle X^r-(b^r-c)\rangle.$

\begin{prop}\label{HCprop}\emph{[H. C. Williams]}\\
Suppose that $c\neq 0$ is  an $r$-th power in $\F_q$. Let
$\theta^r=b^r-c$ with $\theta \in \F_{q^r}$ and $b\in \F_q$ such
that $b^r-c$ is not an $r$-th power in $\F_q$. Then  letting
$\alpha=b-\theta$,
$$
\alpha^\frac{\sum_{j=0}^{r-1}q^j}{r}\in \F_q
$$
is an $r$-th root of $c$.
\end{prop}

\noindent The usual `square and multiply method' (or `double and
add method' if one uses a linear recurrence relation) requires
roughly $\log \frac{\sum_{j=0}^{r-1}q^j}{r}\approx r\log q$  steps
for the evaluation of $\alpha^\frac{\sum_{j=0}^{r-1}q^j}{r}$, and
therefore the complexity of the algorithm of H. C. Williams is
$O(r^3\log q)$ multiplications in $\F_q$.
 H. C. Williams' result can be  expressed in Algorithm
\ref{HCal} using the recurrence relation technique of Section
\ref{recurrence}.

\begin{algorithm}
\caption{H. C. Williams' $r$-th root algorithm
\cite{HC}}\label{HCal}
\begin{description}
\item [Input :] An $r$-th power residue $c$ in $\F_q$
 \item [Output :] $x\in \F_q$ satisfying $x^r=c$
\end{description}
\begin{algorithmic}[1]
 \State{\bf do} {Choose a random $b\in \F_q$} {\bf until} $b^r-c$ is not an $r$-th power residue.
 \State $M\leftarrow \frac{1+q+\cdots +q^{r-1}}{r}$
 \State $A\leftarrow (b,-1,0,...,0)$\ \
    \qquad \qquad \qquad // $A$ is a coefficient vector of  $\alpha=b-\theta$. //
 \State $A\leftarrow$ RecurrenceRelation($A,M$)  \qquad //  $A$ is a coefficient vector of $\alpha^M$. //
 \State $x\leftarrow$ corresponding element of $A$ \qquad //   $x=\alpha^M$ //
 \State {\bf return} $x$
\end{algorithmic}
\end{algorithm}

\noindent Note that $\alpha=b+\theta$ is used in the original
paper \cite{HC}, while our presentation is based on \cite{KS}
where it uses $\alpha=b-\theta$. We followed \cite{KS} because it
is more convenient to deal with general $r$ which is not
necessarily odd prime. For example, if one uses $\alpha=b+\theta$
as in \cite{HC}, then the case of even $r$ (such as $r=2$) cannot
be covered. Detailed explanations will be given in Section
\ref{newalg}.


\subsection{Recurrence relation}\label{recurrence}

Given $\sum_{i=0}^{r-1}a_i \theta^i \in \F_q[\theta]$,
  define $a_i(j)\in \F_q \, (0\leq i\leq r-1,\,\, 1\leq
j)$ as
\begin{align}
\sum_{i=0}^{r-1}a_i(j)\theta^i &= \left(
\sum_{i=0}^{r-1}a_i\theta^i \right)^j.
\end{align}
In particular, one has $a_i(1)=a_i$ for all $0\leq i\leq r-1$.
Then one has
\begin{align}
\sum_{i=0}^{r-1}a_i(m+n)\theta^i&=\left(
\sum_{i=0}^{r-1}a_i(m)\theta^i\right)\left(\sum_{j=0}^{r-1}
a_j(n)\theta^j\right) \notag \\
&=\sum_{l=0}^{r-1}\left(\sum_{j=0}^la_j(m)a_{l-j}(n)\right)
\theta^l+(b^r-c)\sum_{l=0}^{r-2}\left(\sum_{j=l+1}^{r-1}a_j(m)a_{l+r-j}(n)\right)\theta^l,
\notag
\end{align}
which implies
\begin{align}\label{lasteq}
a_l(m+n)&=\sum_{j=0}^la_j(m)a_{l-j}(n)+(b^r-c)\sum_{j=l+1}^{r-1}a_j(m)a_{l+r-j}(n)
\end{align}
for all $0\leq l \leq r-1$. When $l=r-1$, the second summation in
the equation (\ref{lasteq}) does not happen so that one has
$a_{r-1}(m+n)=\sum_{j=0}^{r-1}a_j(m)a_{r-1-j}(n)$. This recurrence
relation is summarized in Algorithm \ref{RecRel}.


\begin{algorithm}
\caption{RecurrenceRelation($A$,$M$)}\label{RecRel}
\begin{description}
\item [Input :] A coefficient vector $A=(a_0,a_1,\cdots,a_{r-1})$
of $a=\sum_{i=0}^{r-1}a_i\theta^i\in\F_q[\theta]$ and
$M\in\mathbb{Z^{+}}$
 \item [Output :] A coefficient vector of $a^M \in\F_q[\theta]$
\end{description}
\begin{algorithmic}[1]
 \State Write $M=\ssum M_i 2^i$ where $M_i\in\{0,1\}$.
 \State $(B_0,B_1,\cdots,B_{r-1})\leftarrow (a_0,a_1,\cdots,a_{r-1})$
 \State {\bf for} $k$ from $\lfloor\log M\rfloor -1$ downto 0 {\bf do}
 \State \qquad
 $(A_0,A_1,\cdots,A_{r-1})\leftarrow(B_0,B_1,\cdots,B_{r-1})$
 \State \qquad {\bf for} $i$ from $0$ to $r-1$ {\bf do}
 \State \qquad \qquad $B_i\leftarrow \sum_{j=0}^{i}A_j A_{i-j}+(b^r-c)\sum_{j=i+1}^{r-1}A_j A_{r+i-j}$
 \State \qquad {\bf if} $M_k=1$ {\bf then}
  \State \qquad \qquad
 $(A_0,A_1,\cdots,A_{r-1})\leftarrow(B_0,B_1,\cdots,B_{r-1})$
  \State \qquad \qquad {\bf for} $i$ from $0$ to $r-1$ {\bf do}
 \State \qquad \qquad \qquad $B_i\leftarrow \sum_{j=0}^{i}A_j a_{i-j}+(b^r-c)\sum_{j=i+1}^{r-1}A_j a_{r+i-j}$
 \State {\bf return} $(B_0,\cdots,B_{r-1})$
\end{algorithmic}
\end{algorithm}

\subsection{An improvement of K. S. Williams and K. Hardy}

Williams and Hardy \cite{KS} improved the algorithm of H. C.
Williams by reducing the loop length to $\log q$ as follows. Write
$\alpha^\frac{\sum_{j=0}^{r-1}q^j}{r}$ (where $\alpha=b-\theta$)
as

\begin{align}\label{e1e2}
\alpha^\frac{\sum_{j=0}^{r-1}q^j}{r}=E_1^\frac{q-1}{r}\cdot E_2,
\end{align}
where
$$
E_1=\alpha^{(q-1)^{r-2}}, \qquad
E_2=\alpha^{\frac{q^r-1}{r(q-1)}-\frac{(q-1)^{r-1}}{r}}.
$$
By noticing that the exponent
${\frac{q^r-1}{r(q-1)}-\frac{(q-1)^{r-1}}{r}}$ of $E_2$ is a
polynomial of $q$ with integer coefficients and using the binomial
theorem, one has the following expression of $E_1$ and $E_2$ as
\begin{align}\label{e1}
E_1=\prod_{i=0}^{r-2}X_i \quad {\rm with}\,\, X_i=(b-\omega^i\theta)^{(-1)^{r-i}\binom{r-2}{i}},
\end{align}
\vspace{-0.5cm}
\begin{align}\label{e2}
E_2=\prod_{i=1}^{r-1}Y_i
                  \quad {\rm with}\,\,Y_i= (b-\omega^{r-i-1}\theta)^\frac{1-(-1)^i\binom{r-1}{i}}{r}.
\end{align}

\noindent Thus we have the following result of Williams and Hardy.

\begin{prop}\emph{[Williams-Hardy]}\\
(1) Under same assumption as in Proposition 1,
$E_1^\frac{q-1}{r}\cdot E_2$ is an $r$-th root of $c$, where
$$
E_1=\alpha^{(q-1)^{r-2}}, \qquad
E_2=\alpha^{\frac{q^r-1}{r(q-1)}-\frac{(q-1)^{r-1}}{r}}.
$$
(2) $E_1$ and $E_2$ can be efficiently computed using the
relations
$$
E_1=\prod_{i=0}^{r-2}(b-\omega^i\theta)^{(-1)^{r-i}\binom{r-2}{i}},
\qquad
E_2=\prod_{i=1}^{r-1}(b-\omega^{r-i-1}\theta)^\frac{1-(-1)^i\binom{r-1}{i}}{r}.
$$
\end{prop}

\begin{algorithm}
\caption{Williams-Hardy $r$-th root algorithm
\cite{KS}}\label{KSal}
\begin{description}
\item [Input :] An $r$-th power residue $c$ in $\F_q$
 \item [Output :] $x\in \F_q$ satisfying $x^r=c$
\end{description}
\begin{algorithmic}[1]
 \State{\bf do} {Choose a random $b\in \F_q$} {\bf until} $b^r-c$ is not an $r$-th power residue.
 \State $\omega\leftarrow (b^r-c)^\frac{q-1}{r}$, where
 $\theta^r=b^r-c$.
 \State $E_1\leftarrow 1,\ E_2\leftarrow 1$
 \State{\bf for} $i$ from 1 to $r-1$ {\bf do}
 \State \quad $X_i\leftarrow (b-\omega^{i-1}\theta)^{(-1)^{r-i+1}\binom{r-2}{i-1}},\ Y_i\leftarrow
 (b-\omega^{r-i-1}\theta)^\frac{1-(-1)^i\binom{r-1}{i}}{r}$
 \State \quad $E_1 \leftarrow E_1\cdot X_i,\ E_2\leftarrow E_2 \cdot
 Y_i$
 \State $A\leftarrow$ coefficient vector of $E_1$
 \State $A\leftarrow$ RecurrenceRelation($A,\frac{q-1}{r}$)
 \State $E_1'\leftarrow$ corresponding element of $A$ in $\F_q[\theta]$
 \State $x \leftarrow E_1'\cdot E_2$
 \State {\bf return} $x$
\end{algorithmic}
\end{algorithm}

\noindent The complexity of computing each of $X_i$ in the
equation (\ref{e1}) is of $O(\log q)+O(r)+O\left(r^2\log
\binom{r-2}{i}\right)$ multiplications in $\F_q$. Hence all $X_i$
can be computed in $O(r\log q+r^4)$ $\F_q$-multiplications. Since
the $O(r)$ multiplications of all $X_i $ ($0\leq i\leq r-2$) in
$\F_{q^r}$ need $O(r^3)$ multiplications in $\F_q$, the total
complexity of computing $E_1$ (as a polynomial of $\theta$ degree
at most $r-1$) is $O(r\log q+r^4)$ $\F_q$-multiplications.
Similarly the complexity of computing $E_2$ is also  $O(r\log
q+r^4)$ $\F_q$-multiplications. For a detailed explanation, see
\cite{KS}. Since the exponentiation $E_1^\frac{q-1}{r}$ (using the
recurrence relation) needs $O(r^2\log\frac{q-1}{r})=O(r^2\log q)$
multiplications in $\F_q$ and since the multiplication of two
elements $E_1^\frac{q-1}r$ and $E_2$ needs $O(r)$ multiplications
in $\F_q$ (because only the constant term of the $\theta$
expansion is needed), the total cost of computing an $r$-th root
of $c$ using the algorithm of K. S. Williams and K. Hardy
\cite{KS} is
  $O(r^2\log q+r^4)$.

\section{Our New $r$-th Root Algorithm}\label{newalg}
In this section, we give an improved version of the Cipolla-Lehmer
type algorithm by generalizing the method of \cite{KS}. Our new
algorithm is applicable for all $r>1$ with $q\equiv 1 \pmod{r}$.
Throughout this section, we assume that $r$ is not necessarily a
prime. Thus $\omega=\theta^{q-1}=\left(b^r-c \right)^\frac{q-1}r$
may not be a primitive $r$-th root of unity even if  $b^r-c$ is
not an $r$-th power in $\F_q$. Consequently a more stronger
condition  is needed for the primitivity of $\omega$.
%
That is, $\omega$ is a primitive $r$-th root of unity if and only
if $\omega^\frac{r}{p}\neq 1$ for every prime $p | r$, which holds
if and only if $(b^r-c)^\frac{q-1}{p}\neq 1$ for every prime
$p|r$. From now on, we will assume that $(b^r-c)^\frac{q-1}{p}\neq
1$ for every prime $p|r$ and therefore $\omega$ is a primitive
$r$-th root of unity.

Let $\alpha \in \F_{q^r}$. Then, by extracting $r$-th roots from
the following simple identity
$$
\alpha^r \left(1\cdot \alpha \cdot \alpha^{1+q}\cdots
\alpha^{1+q+q^2+\cdots +q^{r-2}} \right)^q=\left(1\cdot \alpha
\cdot \alpha^{1+q}\cdots \alpha^{1+q+q^2+\cdots
+q^{r-2}}\right)\alpha^{1+q+\cdots +q^{r-1}},
$$
 one may expect that $ \alpha\left(1\cdot
\alpha \cdot \alpha^{1+q}\cdots \alpha^{1+q+q^2+\cdots +q^{r-2}}
\right)^\frac{q-1}{r}$ equals  $\alpha^\frac{1+q+\cdots
+q^{r-1}}{r}$ up to $r$-th roots of unity. In fact, they are
exactly the same element in $\F_q$ and can be verified as follows;

\begin{align}
\alpha^\frac{1+q+\cdots
+q^{r-1}}{r}&=\alpha^\frac{\sum_{i=0}^{r-1}q^i}{r}=\alpha\cdot\alpha^\frac{(\sum_{i=0}^{r-1}q^i)-r}{r}\\
     &=\alpha\cdot\alpha^\frac{\sum_{i=0}^{r-1}(q^i-1)}{r}
     =\alpha\cdot\alpha^\frac{(q-1)\sum_{i=1}^{r-1}\sum_{j=0}^{i-1}q^j}{r}\\
    &=\alpha\cdot\left(\alpha^{\sum_{i=1}^{r-1}\sum_{j=0}^{i-1}q^j}\right)^\frac{q-1}{r}\\
      &=\alpha\cdot\left(  1\cdot \alpha \cdot \alpha^{1+q}\cdots
\alpha^{1+q+q^2+\cdots +q^{r-2}} \right)^\frac{q-1}{r}.
\end{align}

\begin{prop}\label{MT}\emph{[Main Theorem]}\\
Let $q\equiv 1 \pmod{r}$ with $r>1$ and let $(b^r-c)^\frac{q-1}{p}
\neq 1$ for all prime divisors $p$ of $r$. Then letting
$\alpha=b-\theta$ where $\theta^r=b^r-c$,
$$
\alpha\cdot\left(  1\cdot \alpha \cdot \alpha^{1+q}\cdots
\alpha^{1+q+q^2+\cdots +q^{r-2}} \right)^\frac{q-1}{r}
$$
is an $r$-th root of $c$.
\end{prop}

\noindent Based on the above simple result, we may present a new
$r$-th root algorithm (Algorithm \ref{cubes2}) of complexity
$O(r^2\log q+r^3)$ with given information of the prime factors of
$r$.
 It should be mentioned that our proposed
algorithm is general in the sense that $r$ can  be any (composite)
positive integer $>1$ satisfying $q\equiv 1 \pmod{r}$, while $r$
was assumed to be an odd prime both in \cite{HC} and \cite{KS}.

Both in \cite{HC} and \cite{KS}, $b$ was chosen so that
$\omega=(b^r-c)^\frac{q-1}{r}\neq 1$, and since $r$ is prime,
$\omega$ is automatically a primitive $r$-th root. This property
guarantees the validity of the equation (\ref{Viete}), namely
\begin{equation}\label{keyequation}
(b-\theta)(b-\omega\theta)(b-\omega^2\theta)\cdots
 (b-\omega^{r-1}\theta) =b^r-\theta^r=c.
\end{equation}
However if $r$ is composite, then $\omega=(b^r-c)^\frac{q-1}{r}$
is not a primitive $r$-th root in general. In fact, letting $s>1$
be the least positive integer satisfying $\omega^s=1$, the degree
of the irreducible polynomial of $\theta$ (where $\theta^r=b^r-c$)
is $s$ because
$$\theta^{q^s-1}=(\theta^{q-1})^{q^{s-1}+q^{s-2}+\cdots +q+1}
                =\omega^{q^{s-1}+q^{s-2}+\cdots +q+1}
                =\omega^s, $$
 and one has
\begin{align}
(b-\theta)(b-\omega\theta)\cdots
 (b-\omega^{r-1}\theta) =\{(b-\theta)(b-\omega\theta)\cdots
 (b-\omega^{s-1}\theta)\}^\frac{r}{s}=(b^s-\theta^s)^\frac{r}{s}\neq c
\end{align}
if $s<r$. Therefore the methods of \cite{HC} and \cite{KS} do not
work for a composite $r$ unless one assumes the primitivity of
$\omega$.

\begin{algorithm}[t]
\caption{Our new $r$-th root algorithm  } \label{cubes2}
\begin{description}
\item [Input :] An $r$-th power residue $c$ in $\mathbb{F}_q$
 \item [Output :] $x \in \F_q$
satisfying $x^r=c$
\end{description}
\begin{algorithmic}[1]
 \State {\bf do} Choose a random $b\in \F_q$ {\bf until}
$(b^r-c)^\frac{q-1}{r}$ is a primitive $r$-th root of unity.
 \State $\omega\leftarrow (b^r-c)^\frac{q-1}{r},\ \alpha \leftarrow b-\theta$ where
 $\theta^r=b^r-c$.
  \State  $P\leftarrow \alpha, A \leftarrow \alpha, W\leftarrow 1$
  \State {\bf for} {$i=1$} {\bf to} $r-2$ {\bf do}   \qquad \qquad //$A, P \in \F_q[\theta]$ and $W\in \F_q$//

   \State \quad  $W\leftarrow  W\omega$, $V\leftarrow b-W\theta$
      \,\, \quad //$W=\omega^i, V=b-\omega^i\theta=\alpha^{q^i}$//
   \State \quad  $A\leftarrow AV$, $P\leftarrow PA$ \qquad \,\, \quad //$A=\alpha^{1+q+\cdots+q^{i}}, P=\alpha\cdot\alpha^{1+q}\cdots\alpha^{1+q+\cdots+q^{i}}$//
        \State $B\leftarrow$ coefficient vector of $P$
       \State  $B\leftarrow$ RecurrenceRelation$(B,\frac{q-1}r)$
 \State $P\leftarrow$ corresponding element of $B$ in
 $\F_q[\theta]$
              \State  $x\leftarrow \alpha \cdot P$ \qquad // $x\in \F_q$ //

  \State {\bf return} $x$
\end{algorithmic}
\end{algorithm}

Also, even if one assumes the primitivity of
$\omega=(b^r-c)^\frac{q-1}{r}$, one still has some problems both
in \cite{HC} and \cite{KS}, which will be explained in the
following remarks.

\begin{rem}\label{generalr}
In \cite{HC}, $\alpha=b+\theta$ was used (instead of $b-\theta$)
under the assumption of $\theta^r=c-b^r$ with
$(c-b^r)^\frac{q-1}{r}\neq 1$. If we choose $\alpha=b+\theta$
following \cite{HC}, then we get
\begin{equation}
\begin{split}
(b+\theta)(b+\omega\theta)\cdots
(b+\omega^{r-1}\theta)=b^r-(-\theta)^r=b^r+(-1)^{r+1}\theta^r.
\end{split}
\end{equation}
Therefore if $r$ is odd prime (as was originally assumed in
\cite{HC}), one
 has $b^r+\theta^r=c$ and the $r$-th root algorithm is essentially
 same to the case $\alpha=b-\theta$. However when $r$ is even (for
 example, when
 $r=2$), the original method in \cite{HC} cannot be used
 because $b^r+(-1)^{r+1}\theta^r=b^r-\theta^r\neq c$.
\end{rem}

\begin{rem}
The algorithm in \cite{KS} needs $E_1$ and $E_2$ satisfying
$\alpha^\frac{\sum_{j=0}^{r-1}q^j}{r}=E_1^\frac{q-1}{r}\cdot E_2$.
However for composite $r$, $E_2$ cannot be well-defined in some
cases, since the exponent $\frac{1-(-1)^i\binom{r-1}{i}}{r}$ in
the equation (\ref{e2}) is not an integer in general. That is, the
property $(-1)^i {{r-1}\choose{i}} \equiv 1\pmod{r}$ only holds
when $r$ is prime. Therefore the algorithm in \cite{KS} fails to
give the answer when $r$ is composite such as $r=4,6,9,\cdots$.
(i.e., when $r=4$, one has $E_2=\alpha^{q^2-\frac12 q}$ so the
coefficient $\frac12$ of $q$ in the exponent is not an integer and
one cannot compute $E_2$.) The problem of $E_2$ being undefined is
unavoidable even if one assumes the primitivity of $\omega$.
\end{rem}

\section{Complexity Analysis and Comparison}

\subsection{Complexity analysis}


An initial step of the proposed algorithm requires one to find a
primitive $r$-th root $\omega$ in $\F_q$. When $r$ is prime, one
only needs to find $b$ satisfying
$\omega=(b^r-c)^\frac{q-1}{r}\neq 0, 1$ and the probability that a
random $b$ satisfies the required property is
$\frac{1}{r}+O(q^{-\frac{1}{4}})$ (\cite{KS} pp.480) under the
assumption of $r\leq q^\frac{1}{4}$. When $r$ is composite, one
further needs to check whether $\omega^\frac{r}{p}\neq 1$ for
every prime divisor $p$ of $r$. Since the complexity estimation
$O(r^3\log q)$ in \cite{HC} and $O(r^2\log q+r^4)$ in \cite{KS}
still hold if one assumes that a primitive root
$\omega=(b^r-c)^\frac{q-1}{r}$ is already given, we will also
assume that a primitive root $\omega$ is given in our estimation
for a fair comparison.

At each $i$-th step of the for-loop of our proposed algorithm,
step 5 needs 1 $\F_q$ multiplication. In step 6, the computation
$AV$ needs 1 $\F_{q^r}$ multiplication which, in fact, can be
executed with $2r$ $\F_q$ multiplications because $V=b-\omega^i
\theta$ is linear in $\theta$. The computation $PA$ needs 1
$\F_{q^r}$ multiplication which can be executed with $r^2$ $\F_q$
multiplications. Therefore, at the end of the for-loop, one needs
at most $(r-2)(1+2r+r^2)<(r+1)^3$ $\F_q$ multiplications (of order
$O(r^3)$). Since the exponentiation $P^\frac{q-1}{r}$ (in steps
7-9) needs $O(r^2\log q)$ $\F_q$ multiplications, the total cost
of our proposed algorithm is $O(r^3+r^2\log q)$ multiplications in
$\F_q$. On the other hand, the cost of Algorithm \ref{HCal}
\cite{HC} is $O(r^2\log \frac{q^r-1}{r})=O(r^3\log q)$, and the
cost of Algorithm \ref{KSal} \cite{KS} is $O(r^4+r^2\log q)$ where
$O(r^4)$ comes from the cost of computing $E_1$ and $E_2$ in steps
4-6 of Algorithm \ref{KSal}. The theoretical estimation shows that
our proposed algorithm is better than Algorithm \ref{KSal} as $r$
gets larger.

Finally, when $r=2$, the for-loop can be omitted in our algorithm
so that one only needs to compute $P\cdot P^\frac{q-1}{2}$ which
is exactly same to the original Cipolla-Lehmer algorithm.




\subsection{Implementation results}

Table \ref{table} shows the implementation results using SAGE of
the above mentioned two algorithms and our proposed one. The
implementation was performed on Intel Core i7-4770 3.40GHz with
8GB memory.

\begin{table}[h]
\caption{Running time (in seconds) for $r$-th root algorithms
}\label{table}
\begin{center}
\begin{tabular}{|c||c|c|c|c|c|c|c|}
  \hline
   $r$ & 3 & 4& 43 & 101 & 211  \\
   \hline \hline
 Algorithm \ref{HCal} \cite{HC} & 0.467 & fail & 2026.962 & Interr. & Interr.  \\
  \hline
   Algorithm \ref{KSal} \cite{KS} & 0.254 &fail & 53.849 & 535.043 & 3956.433 \\
    \hline
  Our proposed algorithm  & 0.253 &0.355& 48.359 & 256.601 & 1098.401 \\
  \hline
\end{tabular}
\end{center}
\end{table}

For convenience, we used prime fields $\mathbb F_p$ with size
about $2000$ bits. Average timings of the $r$-th root computations
for 5 different inputs of $r$-th power residue $c \in \mathbb F_p$
are computed for the primes $r=3, 43, 101, 211$. As one can see in
the table, our proposed algorithm performs better than the
algorithms in \cite{HC} and \cite{KS}. The table also shows that
our  algorithm gets dramatically faster than other algorithms as
$r$ gets larger. For example, when $r=101$, our algorithm is
roughly 2 times faster than Algorithm \ref{KSal},  and when
$r=211$, our algorithm is 4 times faster than Algorithm
\ref{KSal}. For $r=101, 211$, the SAGE computation were
interrupted after 3 hours for Algorithm \ref{HCal}.

\section{Conclusions}

We proposed a new Cipolla-Lehmer type algorithm  for $r$-th root
extractions in $\F_q$. Our algorithm has the complexity of
$O(r^3+r^2\log q)$ multiplications in $\F_q$, which improves the
previous results of  $O(r^3\log q)$ in \cite{HC} and of
$O(r^4+r^2\log q)$ in \cite{KS}. Our algorithm is applicable for
any integer $r>1$, whereas the previous algorithms are effective
only for odd prime $r$. Software implementations via SAGE also
show that our proposed algorithm is consistently faster than the
previously proposed algorithms, and  becomes much more effective
as $r$ gets larger.

\end{document}